\documentclass[aps,pre,preprint,longbibliography,byrevtex]{revtex4-1}

\usepackage{amsmath,amsthm, amssymb,pifont}
\usepackage{graphicx,graphics,color,soul, bm}
\usepackage{subfigure, subfig}
\usepackage[usenames,dvipsnames]{pstricks}

\newcommand{\comment}[1]{} % For omitting a passage inside \comment{ }
\begin{document}

\title{Effects of inertia and turbulence on rheological measurements of neutrally-buoyant suspensions }

\author{Esperanza Linares-Guerrero}
\affiliation{Division of Engineering and Applied Science, California Institute of Technology, Pasadena CA 91125, USA}

\author{Melany L. Hunt}
\affiliation{Division of Engineering and Applied Science, California Institute of Technology, Pasadena CA 91125, USA}
\email{hunt@caltech.edu}

\author{R. Zenit}
\affiliation{Instituto de Investigaciones en Materiales, Universidad Nacional Aut\'onoma de M\'exico, Apdo. Postal 70-360, M\'exico Distrito Federal 04510, M\'exico}
\email{zenit@unam.mx}

\begin{abstract}

For low-Reynolds number shear-flows of neutrally-buoyant suspensions, the shear stress is often modeled using an effective viscosity that depends only on the solid fraction. As the Reynolds number ($Re$) is increased and inertia becomes important, the effective viscosity also depends on the Reynolds number itself. The current experiments measure the torque for flows of neutrally-buoyant particles in a coaxial-cylinder rheometer for solid fractions, $\phi$, from 10\% to 50\% and Reynolds numbers based on particle diameter from 2 to 1,000. For experiments for Reynolds of O(10) and  solid fractions less than $30\%$, the effective viscosity increases with Reynolds number, in good agreement with recent numerical simulations found in the literature. At higher solid fractions over the same range of $Re$, the results show a decrease in torque with shear rate. For Reynolds numbers greater than 100 and lower solids concentrations, the effective viscosity continues to increase with Reynolds number. However, based on comparisons with pure fluid measurements the increase in the measured effective viscosity results from the transition to turbulence. The particles augment the turbulence by increasing the magnitude of the measured torques and causing the flow to transition at lower Reynolds numbers. For the highest solid fractions, the measurements show a significant increase in the magnitude of the torques, but the effective viscosity is independent of Reynolds number.

\end{abstract}

%\begin{keywords}
%\end{keywords}

\maketitle

\section{Introduction}
There is an extensive body of work on the rheology of neutrally-buoyant suspensions \citep{Larson:1998,Batchelor:1970,Foss:2000,Stickel:2005}. Typically these studies investigate suspensions in which the  Reynolds number is small, $Re\ll 1$, where $Re$ is defined as $Re=\rho\dot{\gamma}d^2/\mu$, $\rho$ is the fluid density, $\dot{\gamma}$  is the shear rate, $d$ is the particle diameter, and $\mu$ is the fluid viscosity. For low-Reynolds-numbers suspensions, the shear stress, $\tau$, is assumed to depend linearly on  $\dot{\gamma}$ and an effective viscosity, $\mu'$; hence, $\tau=\mu' \dot\gamma$. In this regime, the effective viscosity depends only on the solid fraction, $\phi$, and can be estimated through various semi-empirical relations, such as the Eilers or Krieger-Dougherty relations. For solid fractions greater than approximately 40\%, the effective viscosity may also depend on the shear rate \citep{Stickel:2005, Larson:1998, Foss:2000, Brown:2010}.

As the particle Reynolds number increases, both the inertia of the fluid and solid phases become important \citep{Batchelor:1970, Kulkarni:2008, Verberg:2006}. Recent simulation studies
 calculate the magnitude of the total stress as a function of the Reynolds number and solid fraction \citep{Kulkarni:2008, Yeo:2013, Picano:2013}.
\citet{Kulkarni:2008} used the lattice-Boltzmann method to calculate the particle contribution to the bulk stress for a suspension in a wall-bounded shear flow.  These simulations were performed for particle Reynolds numbers from 0.04 to 16 (note, that Kulkarni and Morris defined a Reynolds number based on particle radius rather than the definition used in this paper) and for solid fractions from 0.05 to 0.3. They computed the effective viscosity in two ways: from the shear stress on the bounding walls and from a computation of the volume-averaged stresses within the bulk of the flow. The two methods produced similar results for the lowest solid fractions and showed about a 10\% difference at the highest solid fraction because of particle slip at the walls and a lower reduced shear rate in the central region of the flow.

\begin{figure}
\centering
\includegraphics[width=0.9\textwidth]{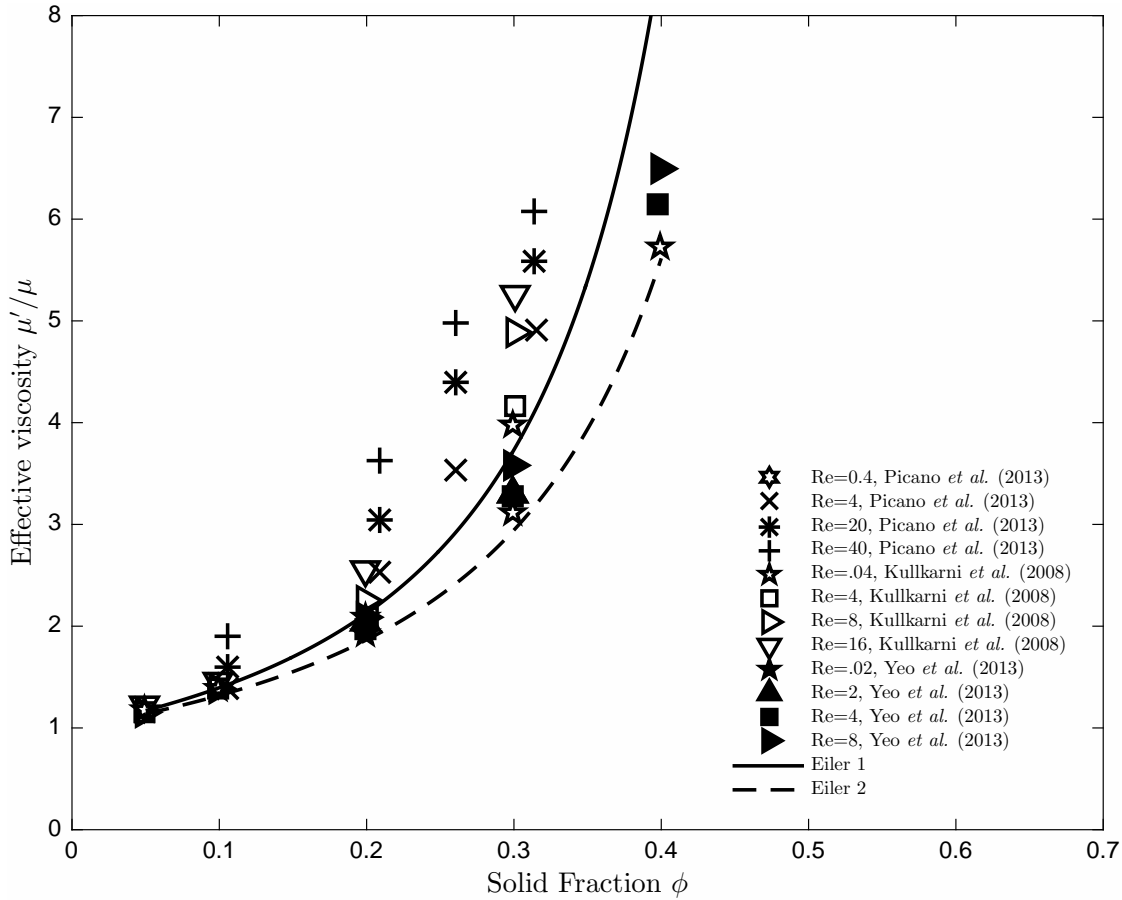}% Images in 100% size
\caption{Effective viscosity for inertial suspensions as a function of solid fraction from numerical simulations made by \citet{Kulkarni:2008,Picano:2013,Yeo:2013}. Continuous and dashed lines correspond to Eilers relation with $a=1.5$ and $\phi_m=0.58$ and $a=1.25$ and $\phi_m=0.63$ respectively.}
\label{fig1}
\end{figure}

The effective viscosity results of \citet{Kulkarni:2008} are found in Fig. \ref{fig1} as a function of solid fraction and Reynolds number.  At the lowest solid fractions ($\phi<0.2$) and Reynolds numbers ($Re <4$), the dependence of $\mu '$ on solid fraction matches the Eilers relation, $\mu'=\mu[(1+a\phi)/(1-\phi/\phi_m)]^2$, using $a=1.5$ and $\phi_m=0.58$ as assumed by \citet{Kulkarni:2008} (identified as Eilers 1 in Fig. \ref{fig1}). At $\phi=0.2$ the effective viscosity shows an increase with Reynolds number with the effective viscosity at $Re=16$ twenty percent higher than found at $Re=0.04$. At $\phi=0.3$, their simulations showed a slight shear thinning (approximate $6\%$ drop) for $Re<0.4$, and an increase of approximately $33\%$ at $Re=16$ compared to the values at the $Re=0.4$.  The authors stated that for $\phi=0.3$, the slight drop in the effective viscosity and then subsequent rise resulted from ordering of the microstructure in the flow direction followed by an increase in the symmetric first moment \citep{Batchelor:1970} for increasing Reynolds numbers. Over the range of conditions considered, the authors found that the contributions from the acceleration and Reynolds stresses were small compared with the contribution from this term \citep{Kulkarni:2008}.

\citet{Yeo:2013} considered a uniform shear flow rather than a wall-bounded flow. They use a lubrication-corrected, force-coupling method to calculate the stresses and the self-diffusion coefficient for $Re$ from 0.02 to 8 and solid fractions from 0.2 to 0.4. In their simulations, they did not include the Reynolds stresses in the calculation of the effective viscosity. As shown in Fig. \ref{fig1}, the effective viscosity from their simulations increased as the Reynolds number increased above one; however, at $\phi=20\%$ the effective viscosity is approximately 7\% lower than \citet{Kulkarni:2008}  and at $\phi=30\%$ it is 37\% lower at the highest Reynolds number studies ($Re=8$); in their paper, they compared with the Eilers relation using $a=1.25$ and $\phi_m=0.63$ (Eilers 2 in Fig. \ref{fig1}).  Although the Reynolds stresses were not included in the calculation, they estimate that the magnitude of such stresses are an order of magnitude smaller than the particle stresses. They also argue that using a pure shear flow, as opposed to the wall-bounded flow used by \citet{Kulkarni:2008}, may contribute to the differences in the calculations of $\mu '$, especially at higher solid fractions.

\citet{Picano:2013} also computed the effective viscosity for Reynolds numbers from 0.4 to 40 and solid fractions from 0.11 to 0.31; their results are also found in Fig. \ref{fig1}. They used an immersed boundary method from \citet{Breugem:2012} to compute the stresses in a wall-bounded suspension, similar to \citet{Kulkarni:2008}. Their results are up to 15\% greater than those found by \citet{Kulkarni:2008}. Note that to compute the effective viscosity, they used the shear stress at the wall and divided by the average shear rate across the channel; as shown by \citet{Kulkarni:2008} for $\phi$=0.3, this calculation gives a higher effective viscosity than found by computing the effective viscosity from the particle contribution to the volume-averaged bulk stress within the flow. This same numerical method has recently been used to examine laminar and turbulent flow of a suspension in a channel for a channel Reynolds numbers up to 5,000 and in which a particle Reynolds number is greater than 100 \citep{Picano:2015}.

An additional numerical study by \citet{Trulsson:2012} considered the transition from viscous to \emph{inertial} regime in dense two-dimensional suspensions using a discrete element model coupled to a fluid solver. Their results show a transition from linear to quadratic dependence on the shear rate at volume fractions that are close to the jamming transition. Although the authors use the term ``inertial suspension" they assume a Stokes drag to couple the particles and fluid and note that the Reynolds number is small in all simulations.

Besides these recent numerical studies, several experimental studies have considered the effect of inertia on suspension rheology \citep{Bagnold:1954,Savage:1983,Hanes:1985, Prasad:1995, Dijksmann:2010}. However, there are important distinctions between the experimental and numerical studies that complicate the comparisons of the results. In simulations, the density of the particles can be specified to equal that of the fluid. In experimental work, the particle and fluid densities may be difficult to match over all experimental conditions because of the temperature dependence of the densities. As discussed by \citet{Acrivos:1994}, \citet{Fall:2009} and \citet{Dijksmann:2010} any subsequent settling or floating of the particles can have a significant effect on the stress measurements. In addition for studies involving flow inertia, a simple-shear flow is difficult to achieve because of macro-scale flow transitions that may occur as a result of the experimental design \citep{Hunt:2002}.

For this study, the focus is on neutrally-buoyant suspensions with inertia. In the 1950s, Bagnold conducted experiments using 1-mm wax particles in water ($Re$ ranged 20 to 400 and $\phi$ from 0.13 to 0.62) and the same particles in a more viscous water-glycerol-ethanol mixture ($Re$ from 2 to 15 for $\phi=0.55$). From the experimental measurements, Bagnold concluded that the flow transitioned from a linear, `macroviscous' regime in which the stresses varied linearly with shear rate to non-linear, inertial regime in which the stresses varied quadratically on the shear rate.  The experimental measurements, however, were conducted in a smooth-walled concentric cylinder rheometer with a rotating outer cylinder, rotating end plates, and stationary inner cylinder. As shown by \citet{Hunt:2002}, the gap Reynolds number associated with the rheometer design (here the Reynolds number is based on the speed of the rotating wall and the gap size of the annulus, $Re_b=\rho\omega r_ob/\mu$, where $\omega$ is the rotational speed, $r_o$ is the outer cylinder radius and $b$ is the shear gap width) ranged from 800 to 28,000; the height of the rheometer, $H$, was small relative to the gap, $H/b=4.6$. Hence for many of the experiments, the Reynolds number was high enough that the flow was no longer a simple shear flow but contained counter-rotating vortices at the end walls that contributed substantially to the torque and stress measurements. \citet{Hunt:2002} concluded that the variation in shear stresses from a linear regime to a non-linear regime resulted from the development of the vortical structures and not from particle-to-particle collisions described by Bagnold.

Following \citet{Bagnold:1954}, \citet{Savage:1983} used neutrally-buoyant 1-mm polystyrene spheres in a co-axial cylindrical rheometer with a rotating inner cylinder. Because their experiments involved gap Reynolds numbers beyond the critical Reynolds number, their rheological measurements were strongly affected by the secondary flows that resulted from the centrifugal instability due to the rotation of the inner cylinder.

As Reynolds numbers are increased in a suspension flow, turbulent fluctuations may contribute to the momentum transport. A study by \citet{Gore:1989} compiled experimental measurements of the change in turbulent fluctuations for multiphase flows (liquid-solid, gas-solid, gas-liquid, and liquid-gas) relative to single-phase flows in pipes or free jets.  Their analysis across 15 different data sets showed for flows in which the particle diameter was greater than 10\% of turbulence length scale ($d/l_t \geq 0.1$, where $l_t$ is the turbulence length scale), such as the size of the most energetic eddy, the turbulent intensity of the carrier phase was increased. They concluded that particles larger than the most energetic eddy do not follow the turbulent motions and the relative motion produces wakes that enhance the turbulence.  For pipe flows, they used a turbulence length scale $l_t\approx0.1D$, where $D$ is the pipe diameter; hence, their criteria suggest that particles with diameters greater than 1\% of the pipe diameter enhance the turbulent fluctuations. Work by \citet{Tanaka:2008} argues that the concentration, density ratio, and Reynolds number are also important in determining the augmentation or attenuation of the turbulent fluctuations. As a note, these studies did not include neutrally-buoyant particles and the studies focused on low solid fractions.

Another interesting aspect of particulate flows is the appearance of fluid velocity fluctuations (i.e. Reynolds stresses) resulting from the presence of a dispersed phase and not from the instability of the flow resulting from inertia. Both particulate and bubbly flows have been found to exhibit large Reynolds-like stresses at low Reynolds numbers that increase with the particle volume fraction \citep{Cartellier:2001,Martinez:2007,Mendez:2013}. Since the origin of these fluctuations is different from that of single phase turbulent flows, these flows are often called pseudo-turbulent. It has been shown that the velocity fluctuations are a result of the distribution of particles in space rather than due to the nature of the flow around the particles \citep{risso:2011}. It is also important to point out that self-diffusion in particulate flows, a subject that has been widely studied \citep{Breedveld:1998}, is closely related to these turbulent-like fluctuations.  These particle-induced fluctuations and an enhanced diffusivity are expected to influence the transition to ordinary turbulence but this aspect has not been addressed in detail in the literature. It should be noted that finite-inertia suspensions, in particular wall-bounded flows, are not necessarily homogeneous; therefore their analysis and understanding are more complex.  For instance, \citet{Verberg:2006} and \citet{Kulkarni:2008} have reported on the effects of an inhomogeneous shear rate within bounded flows.

Neutrally-buoyant particles were considered by \citet{Matas:2003} in their experimental study of the effects of particles on the critical Reynolds number for flow in a pipe. Their results, based on pressure fluctuations in the flow, showed that critical Reynolds number depends on both the ratio of the pipe diameter to the particle diameter ($D/d$) and on the solid fraction. For particles with $D/d \geq65$, the critical Reynolds number increased for all solid fractions examined (up to $\phi=0.35$) and was independent of particle size; this delay in turbulent transition results from the increase in effective viscosity caused by the suspended particles.  For larger particles in which $D/d \leq65$, the critical Reynolds number either increased or decreased. For the largest particles considered ($D/d\approx10$), the critical Reynolds number dropped from approximately 2,000 to 1,000 at $\phi=0.07$; the authors suggest that the presence of the particles caused the flow to transition at a lower Reynolds number. Beyond $\phi=0.07$, the critical Reynolds number increased reaching approximately 2,700 at $\phi=0.35$.

The current study follows the work of \citet{Koos:2012} and uses the same experimental facility, which is described in the next section. In that study, the authors showed that for the range of Reynolds numbers, solid fractions, and particles tested ($Re$ from 20 to 800; $\phi$ from 0.07 to 0.6; three different types of particles) the effective viscosity had little dependence on the Reynolds number. However, the magnitude of the effective viscosity was substantially larger than predicted by an Eilers or Krieger-Dougherty model. In most of the experiments, the rheometer walls were smooth and the authors found substantial slip along with the walls. Hence, that work included some additional measurements for one type of particle with roughened walls. The roughness resulted in a further increase in the effective viscosity, presumably due to the increased agitation of the particles; however, the variation in the effective viscosity with Reynolds number only showed a slight increase at the lowest solid fractions. The experiments in this paper focus on measurements with roughened walls and neutrally-buoyant particles. This work expands the conditions considered in \citet{Koos:2012} to include a wider range of  Reynolds numbers ($Re$ ranges from 2 to 1,000); in addition, this work reconsiders some of the experimental data found in \citet{Koos:2012}.

\section{Experimental Rheometer and Pure Fluid Studies}\label{sec2}

\begin{figure}
\centering
{\includegraphics[width=\textwidth]{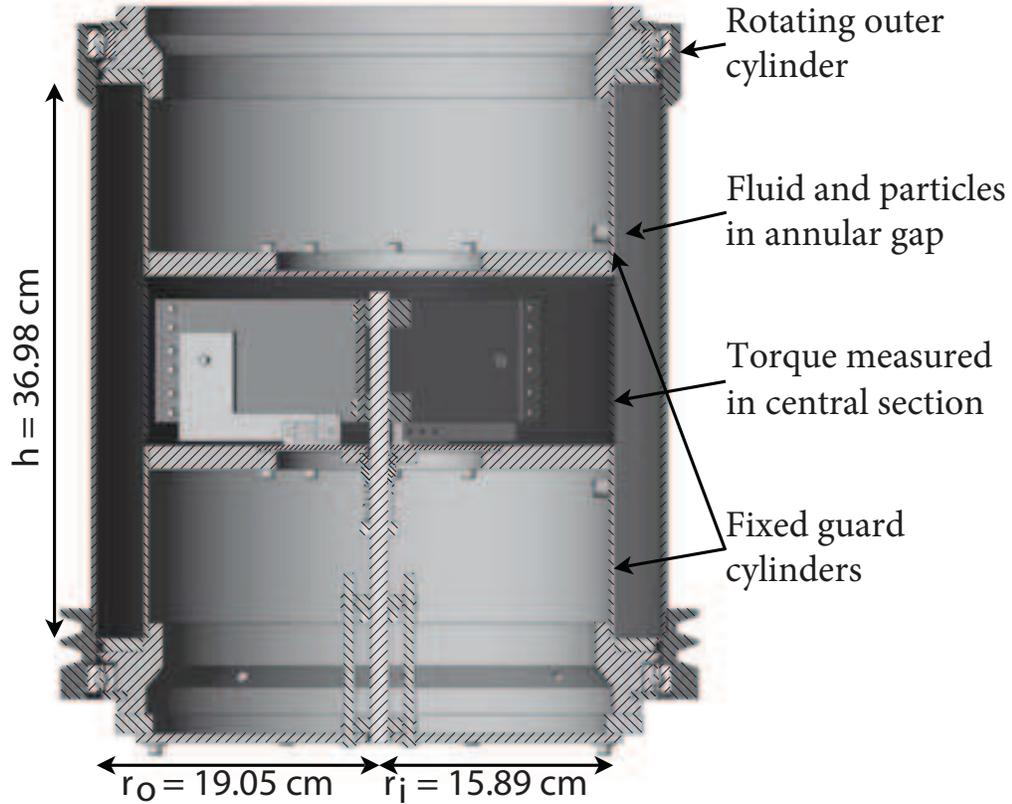}}% Images in 100% size
\caption{Schematic of the coaxial cylinder rheometer.}
\label{fig2}
\end{figure}

This study uses the coaxial-cylinder rheometer, as shown in Fig. \ref{fig2} and described in detail by \citet{Koos:2012}. The inner cylinder is stationary and the outer cylinder is rotated by a belt connected to a motor; the maximum angular speed is $\omega= 15 s^{-1}$. To allow the torque measurements to be made in a region of the flow away from the corners where the flow is not a simple shear flow, the inner cylinder consists of three sections: the rigid top and bottom guard cylinders (each of height 12.7 cm) and a central, floating cylinder ($H=11.22$ cm).  The floating or test cylinder is supported by a central axle and its rotation is constrained by a calibrated spring to allow measurement of the torque or shear stress. To facilitate rotation, the cylinder is separated from the upper and lower guard cylinders by knife-edge gaps. The torque on the cylinder is computed from a measurement of the rotation of the central cylinder, the spring constant and the cylinder radius. The springs were calibrated prior to each experimental run. The spring calibration was done with the rheometer running to ensure similar experimental conditions. Springs with different stiffness are used to allow a range of torques ($M$) between $1.3\times 10^{-3}\leq M \leq2.7$ Nm to be measured. In reviewing the data from \citet{Koos:2012} it was noted that spring calibration was carried out just for pure fluid measurements; for the subsequent experiments with particles the torque measurements were constrained through the origin. Hence the analysis assumed that the relationship between torque and shear rate was linear. The slope of the linear fit was used by \cite{Koos:2012} to find the effective viscosity of the mixture. Therefore, any error in the $y$-intercept in the flow curve (shear stress versus shear rate) can affect the value of the averaged shear stress. As a note, the intercept could have a non-zero value, similar to a yield stress, if the particles had settled because of a density mismatch between the fluid and the particles. In section \ref{discussion}, some of the data is reconsidered without the zeroing of the calibration.

The current experiments are conducted using polystyrene elliptical cylinders with major axis length 2.92 mm and minor axis length 2.08 mm; the particles have smooth walls but are rough cut with average length of 3.99 mm. An equivalent sphere diameter is 3.34 mm. These particles were also used in \citet{Koos:2012} along with spheres and spheroids. Unlike the particle shape dependence found in low Re regime rheological measurements \citep{Mallavajula:2013, Audus:2015}, the measurements from \citet{Koos:2012} did not show a difference with particle geometry.

In \citet{Koos:2012} the random loose-pack and close-pack volume fractions were reported as 0.553 and 0.663. In the current work the rheological measurements are performed with walls roughened, which involved gluing these same particles to thin sheets and then attaching the sheets to the inner and outer cylinders. Without roughness, the inner radius of the annulus, $r_i$, is 15.89 cm, the outer radius of the annulus, $r_o$, is 19.05 cm, and the width of the annulus between the cylinders, $b$, is 3.16 cm. With roughness, these averaged dimensions are $r_i=16.22$ cm, $r_o=18.72$ cm, and $b=2.49$ cm. Note that the device used in this investigation has a gap of approximately 7.5 particle diameters. Clearly, some local inhomogeneities are expected (as discussed by \citealt{Koos:2012}), resulting from this finite gap size. To simplify the analysis, no attempt to correct for this effect was made, which would require a determination of the particle distribution within the gap. The density of the polystyrene particles is 1050 $kg/m^3$, which are the lowest-density particles used by \citet{Koos:2012}.

For the suspension experiments two different liquids were used. To match the density of the particles, the liquids were either a mixture of approximately 79\% water and 21\% glycerol (low viscosity fluid) or a mixture of 58\% ethanol and 42\% glycerol (high viscosity fluid). For each experiment, the fluid density and temperature were measured and used to determine the fluid viscosity. At 22$^o$C, the viscosity of the ethanol-glycerol mixture was 0.030 Pa s and 0.0018 Pa s for the glycerol-water mixture. The differences in densities between the particles and fluid were less than 1\%. Although the density differences were small, some settling or flotation of the particles could be observed due to temperature changes, which was similar to that observed in the experiments by \citet{Acrivos:1994}.

Prior to the suspension studies, tests with liquids without particles were run to test the experimental method. These measurements were conducted considering four different fluids. For all cases, both for the suspensions and pure liquid tests, each measurement was repeated at least five times. In this manner, the standard deviation of each experiment could be calculated to assess the uncertainty of the measurement. For each experimental condition ( a combination of fluid and particle concentration), measurements were obtained by gradually increasing the angular speed of the setup. In some cases, for the same nominal conditions, the experiments were repeated on different days and considering different springs to ensure repeatability of the results.

For a single-phase viscous fluid in annular shear flow with outer rotating wall, the torque, $M_{lam}$, is computed as:
\begin{equation}
M_{lam}=2\pi\mu\dot{\gamma}Hr_i^2.
\end{equation}
where the shear rate, $\dot{\gamma}$, is computed from the rotational speed and annulus geometry, $\dot{\gamma} = 2 \omega r_o^2 /( r_o^2 - r_i^2)$. In \citet{Koos:2012}, pure-fluid torque measurements were also conducted for a smooth-walled rheometer. Those measurements showed laminar behavior up to $Re_b=6,000$; the torques were within 20\% of the laminar-flow values with the highest deviation occurring at the lowest shear rates. For flow between concentric cylinders with rotation of the inner cylinder, there is a complex set of transitions that depend on the Reynolds number of the inner cylinder and the geometry (see \citealt{Taylor:1936a,Taylor:1936b, Coles:1965, Swinney:1978}). However, for a flow in which only the outer cylinder rotates, the flow is ``azimuthal laminar flow with weak Ekman vortices'' \citep{Andereck:1986} below a critical Reynolds number. The Ekman vortices arise because of the end walls; the strength of these vortices depends on the geometry and whether the end walls are fixed or rotate \citep{Coles:1965, Vanatta:1966,Andereck:1986}. For a rotating outer cylinder, the early work by \citet{Taylor:1936a} and \citet{Wendt:1933} used torque measurements to determine the critical Reynolds number at which transition occurs for different radii of the inner and outer cylinders.

Of particular interest for the present study are the papers by \citet{Coles:1965} and \citet{Vanatta:1966}, which also involved flow between rotating cylinders. Their work does not include torque measurements but does examine the transition from laminar to turbulent flow and includes Reynolds numbers over the range found in the current experiments (this range is larger than that found in \citealt{Andereck:1986}.)  They found that for flows with a rotating outer cylinder there was a range of Reynolds numbers in which the flow could be either laminar or turbulent depending on the process for starting the flow and on the height of the annular region relative to the gap. If the Reynolds number was beyond the critical value, the rotational speed had to be reduced to a fraction of the critical speed to return the flow to a laminar state.

In the more recent torque measurements by \citet{Ravelet:2010}, the laminar to turbulent transition is described as ``very sharp''. For these experiments, the height was considerably larger than the gap, $H/b=22$. The critical Reynolds number found in the study of \citet{Ravelet:2010}  was approximately $4000$ for $b/r_o=0.083$, which is consistent with the critical Reynolds number based on Taylor's data. \citet{Ravelet:2010} also measured the circumferential velocity distribution for supercritical speeds and found steep velocity gradients at the inner and outer walls and a flattened central region as compared with the laminar flow profile.  With rotation of the outer wall (their end walls also rotated), they did not observe any large-scale turbulent structures within the flow.

For the smoothed-wall rheometer, the critical Reynolds number can be estimated based on the ratio of the gap width to outer radius ratio, $b/r_o$ from the work by \citet{Taylor:1936a}. For $b/r_o=0.17$, the critical Reynolds number is estimated as $Re_c=1.6 \times 10^4$. With rough walls, the gap width and outer radius are smaller ($b/r_o =0.13$), which results in a lower critical Reynolds number of $1\times10^4$.

\begin{figure}
\centering
{\includegraphics[width=0.9\textwidth]{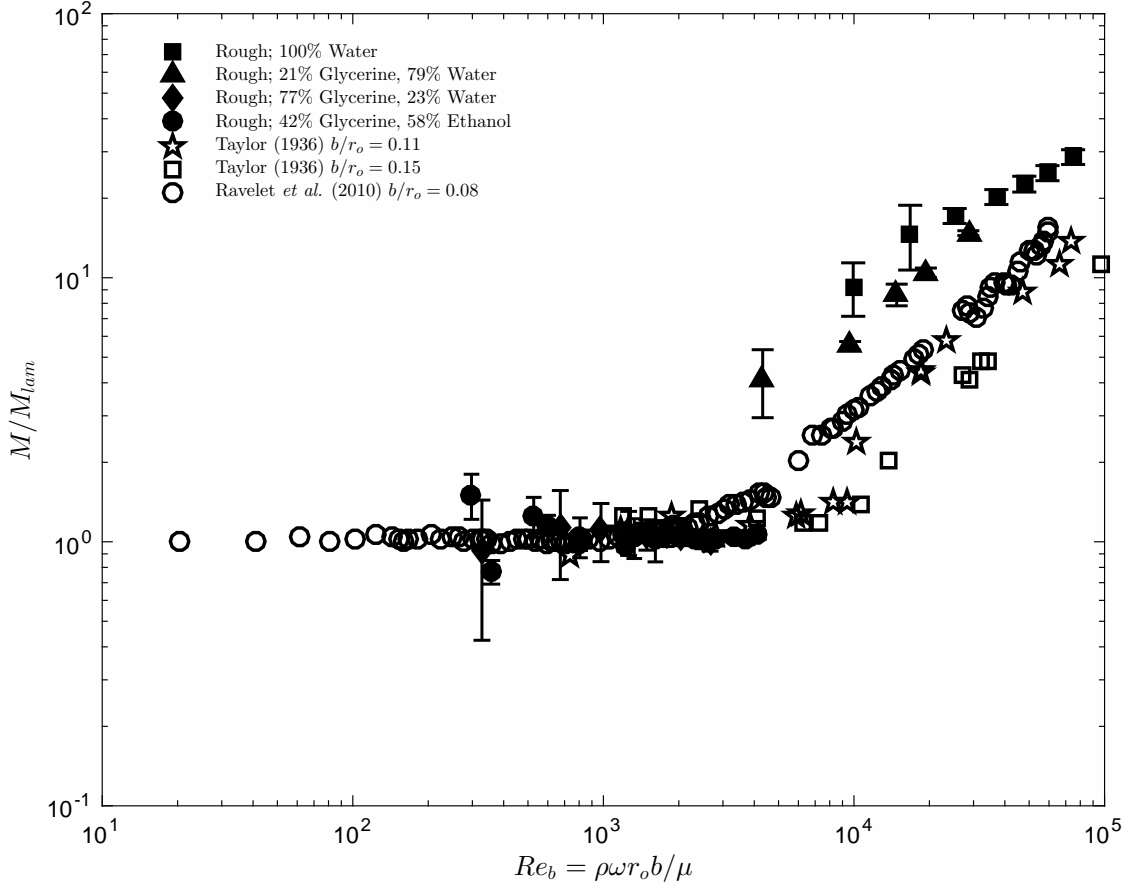}}% Images in 100% size
\caption{Measured torque normalized by the theoretical pure fluid torque considering laminar flow, as a function of  gap Reynolds number for different fluids. Open symbols correspond to the data from \citet{Taylor:1936a} and \citet{Ravelet:2010}.}
\label{fig3}
\end{figure}

Figure \ref{fig3} presents the measured torque for pure fluid, $M$, normalized by $M_{lam}$ for the current work and the data from \citet{Taylor:1936a} and \citet{Ravelet:2010}; the data show a transition around $Re_b =4\times10^3$, lower than the critical Reynolds number based on the work of \citet{Taylor:1936a}. However, the current experiments involve rough-walled cylinders, which appear to cause a transition to turbulent flow at a lower Reynolds number.  This finding contrasts with a study by \citet{Lee:2009} using a rotating inner cylinder and stationary outer cylinder with axial slits; in that study the critical Reynolds number at which the laminar instability first occurred did not change although the transition to turbulence was accelerated.

As found in Fig. \ref{fig3}, one data set from \citet{Taylor:1936a} (for $b/r_o=0.15$) shows a relatively smooth transition, while the second data set (for $b/r_o=0.11$) shows a sharp transition in which the torque increases by a factor of 2. Although not shown, the experiments by \citet{Wendt:1933} also show a sharp transition (this data can also be found in the paper by \citealt{Hunt:2002}).

In addition, the recent experiments by \citet{Ravelet:2010} also show a gradual increase in torque starting around $Re=1000$ followed by a significant jump at $Re\approx 5000$. The variations in these measurements is probably a result in the differences in the height of the annulus relative to the gap size and the method of increasing or decreasing the rotational speed to arrive at the final state \citep{Coles:1965, Vanatta:1966}.

With regard to the current data, Fig. \ref{fig3} shows that the normalized torque does not increase before $Re\approx4000$. At this Reynolds number, the torque ratio shows a sharp increase. The difference between the current data and earlier studies involves the use of the central and guard cylinders; with this experimental design the torque measurements do not include the contributions from the end-wall regions; instead the torque shows a sudden change in magnitude that corresponds with a fully-turbulent flow. Beyond the critical Reynolds number, the normalized torque is greater than that found in the two sets of experiments by \citet{Taylor:1936a} for $b/r_o=0.115$ and 0.148 and in the experiments from \citet{Ravelet:2010} for $b/r_o=0.083$.  However, an increase in magnitude of torque because of wall roughness is consistent with the earlier studies in which the torque was measured for flow in an annulus with an inner rotating cylinder \citep{Cadot:1997,Berg:2003} and with studies involving turbulent flow along a roughened plate or in a tube.

For flow between coaxial cylinders with either one or both cylinders rotating, the dimensionless torque is often expressed as
\begin{equation}
G=ARe_b^\alpha
\end{equation}
where $G=T/(2 \pi H \mu^2 / \rho)$ and $A$ is a constant. For laminar flow, the value of $A$ is computed as $A= 2 r_o r_i^2/[(r_o-r_i)^2 (r_o+r_i)]$ and $\alpha =1$. The current data for $Re_b >4,000$ is best represented using $\alpha \approx1.7$. The study by \citet{Ravelet:2010} for an outer rotating cylinder found $\alpha$ changes from 1 to approximately 1.75 as the Reynolds number increased from 1,000 to 5,000 with a critical Reynolds number of approximately 2,000. In the work by \citet{Berg:2003} for turbulent flow in Taylor-Couette flow with a rotating inner cylinder, they found that the value of the exponent $\alpha$ was equal to 2 when both walls were rough; for smooth walls, $\alpha=1.67$, and for one smooth and one rough wall the value fell between $\alpha=1.8$ to 1.9 for Reynolds numbers (based on the speed of the inner wall) up to $10^6$.

\section{Experiments with Neutrally-Buoyant Particles}

\begin{figure}
\centering
{\includegraphics[width=0.9\textwidth]{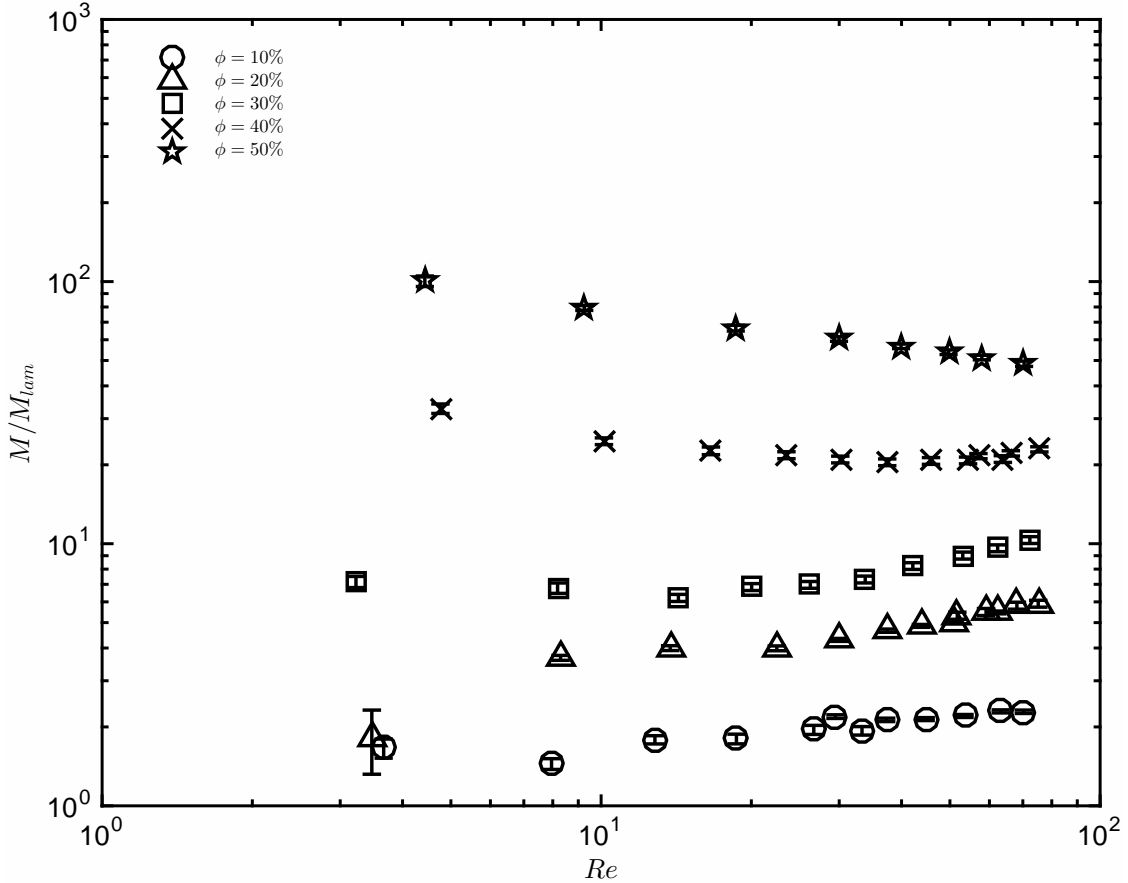}}% Images in 100% size
\caption{Measured torque normalized by the theoretical pure fluid torque considering laminar flow, as a function of Reynolds number for polystyrene particles immersed in a glycerol-ethanol solution (high viscosity fluid.)}
\label{fig4}
\end{figure}

\begin{figure}
\centering
{\includegraphics[width=0.9\textwidth]{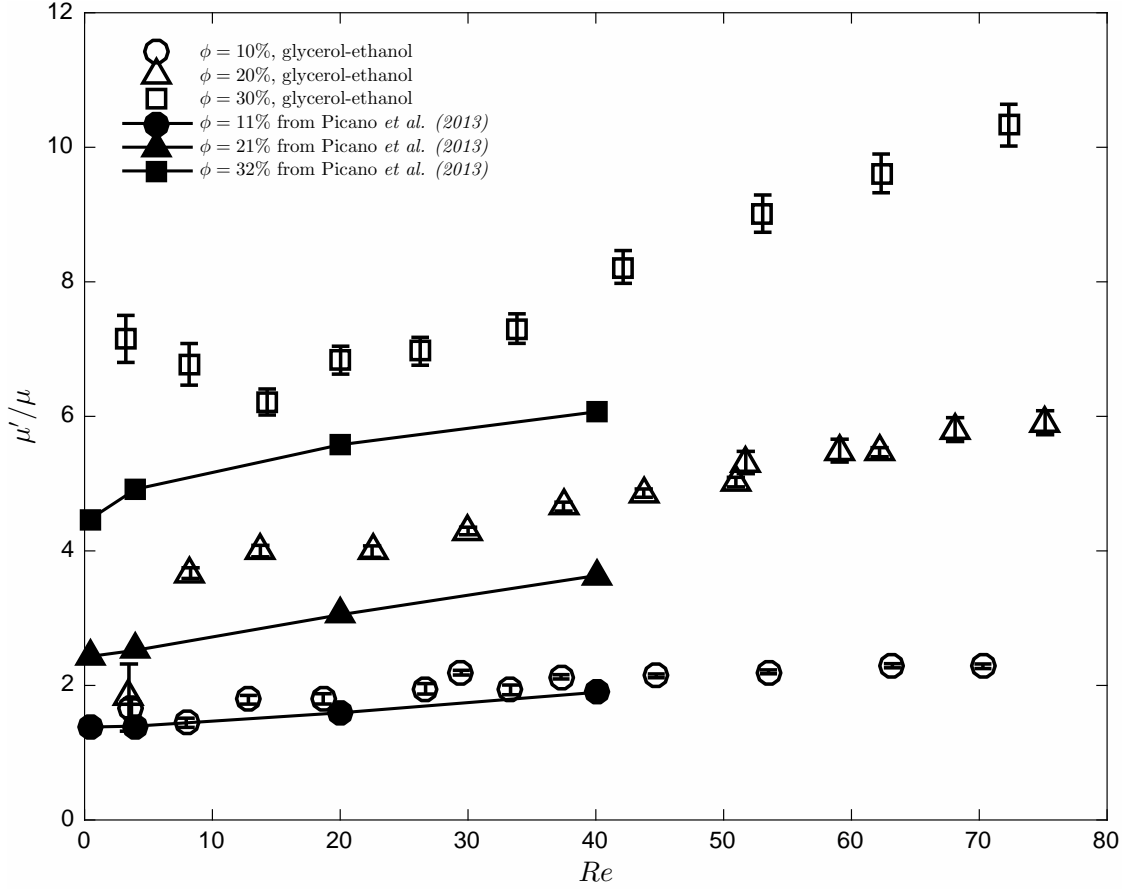}}% Images in 100% size
\caption{Normalized torque (experiments) and effective viscosity (numerical results) as a function of Reynolds number. The experiments correspond to the polystyrene particles immersed in a high-viscosity fluid for $\phi=0.1$, 0.2, and 0.3. The simulations results are from \citet{Picano:2013}.}
\label{fig5}
\end{figure}

Figure \ref{fig4} presents the torque measurements normalized by the torque for laminar flow ($M/M_{lam}$) for the polystyrene particles in the glycerol-ethanol mixture for $\phi$ from 0.1 to 0.5. Figure \ref{fig5} shows on a linear scale the same data for $\phi=0.1$, 0.2, and 0.3 along with the simulations of \citet{Picano:2013}; the data from \citet{Kulkarni:2008} and from \citet{Yeo:2013}  are not shown because of the lower Reynolds numbers considered in those studies. For $\phi=0.1$ the normalized torque, or equivalently the effective viscosity, is approximately constant for $Re$ from 3 to approximately 18 and then increases for higher Reynolds numbers. As shown in Fig. \ref{fig5}, the experimental data for $\phi=0.1$ compares well with the simulations from \citet{Picano:2013}.  For $\phi=0.2$, the normalized torques increase over the range of Reynolds numbers examined.  Except for the experimental data point at $Re=2.5$, the experimental measurements are approximately 40\% higher than the values from the \citet{Picano:2013} simulations. For $\phi=0.3$, the experimental measurements show a slight shear thinning for $Re<10$ followed by a subsequent increase for $Re$ from 15 to 60. At $Re=40$, the experimental measurements are approximately 50\% higher than the simulations. Possible reasons for the differences are discussed in the next section.

For the highest solid fractions, $\phi=0.4$ and $\phi=0.5$, the normalized torques are found in Fig. \ref{fig4} and are distinct from the behavior found at lower solid fractions. For $\phi=0.4$, the normalized torque shows a shear thinning behavior until around $Re=20$, followed by shear thickening. For $\phi=0.5$, the normalized torque decreases over the entire range of Reynolds numbers.

\begin{figure}
\centering
{\includegraphics[width=0.9\textwidth]{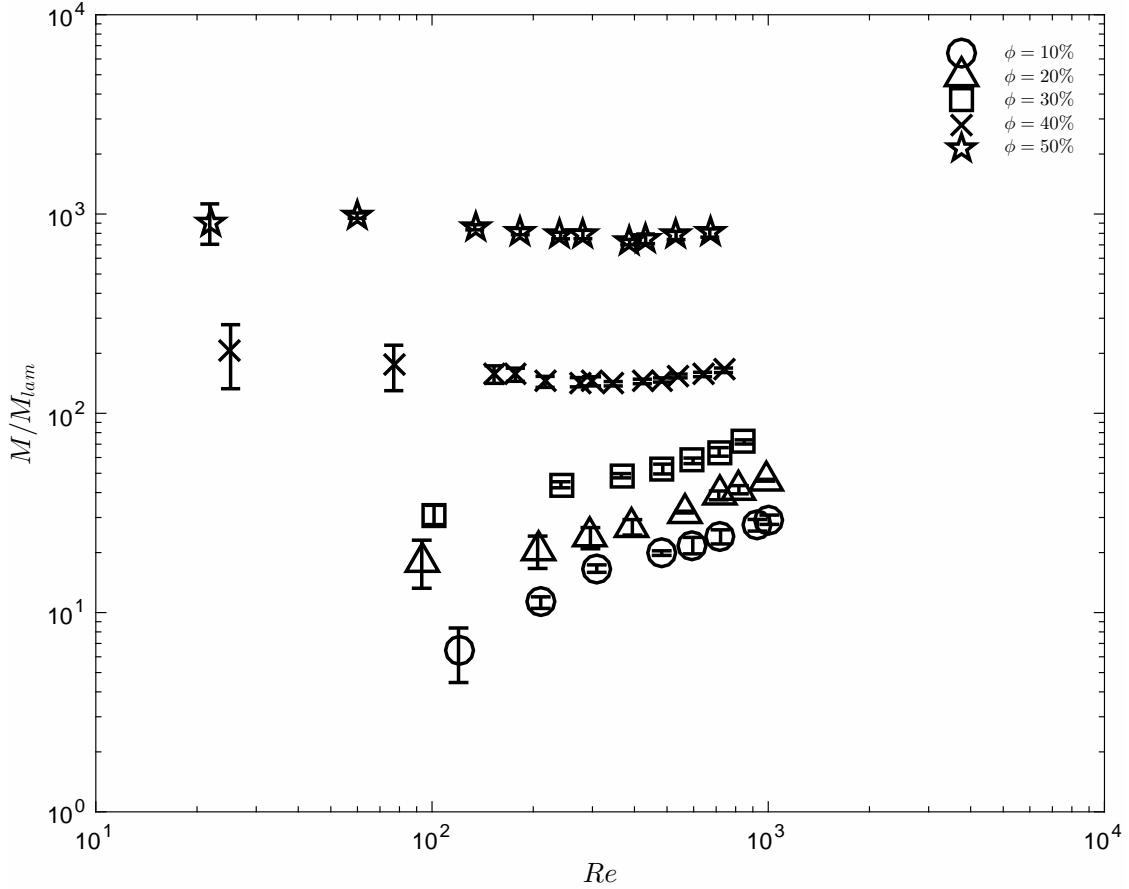}}% Images in 100% size
\caption{Normalized torque as a function of Reynolds number for polystyrene particles immersed in the glycerol-water solution (low-viscosity fluid.)}
\label{fig6}
\end{figure}

Figure \ref{fig6} shows the measurements of the normalized torque ($M/M_{lam}$) for the same particles in the glycerol-water mixture resulting in Reynolds numbers greater than 100. For $\phi=0.1$, 0.2, and 0.3, the normalized torques increase with Reynolds number. For $\phi=0.4$ and $\phi=0.5$, the normalized torques show little variation with Reynolds number.

\begin{figure}
\centering
{\includegraphics[width=0.9\textwidth]{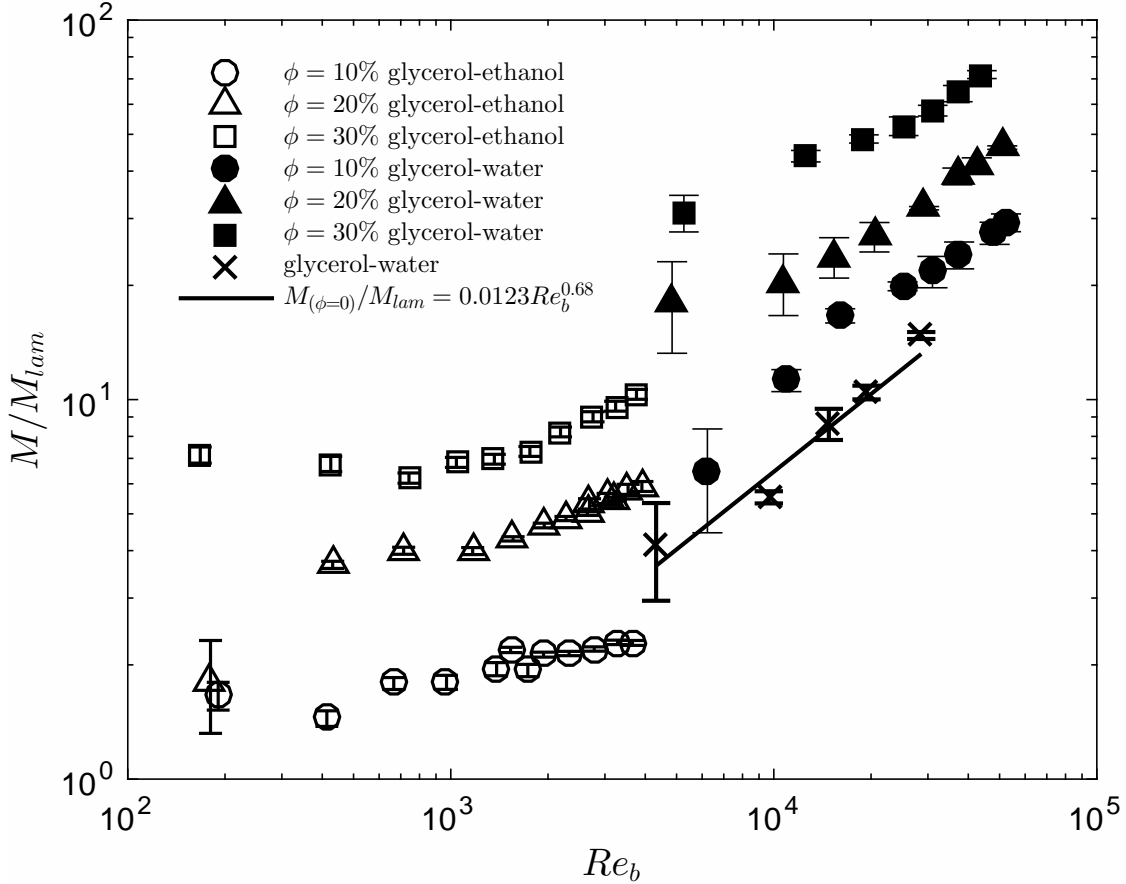}}% Images in 100% size
\caption{Normalized torque as a function of gap Reynolds number for polystyrene particles immersed in the high- and low-viscosity fluids, respectively.}
\label{fig7}
\end{figure}

Figure \ref{fig7} shows both sets of data for $\phi= 0.1$, 0.2, and 0.3 as a function of the gap Reynolds number, $Re_b$. The results using the glycerol-water mixture have torques that are higher than might be anticipated if the torques measured using the glycerol-ethanol mixture are extrapolated to higher Reynolds numbers. As a note the rheometer was run over its full range of speeds; unfortunately there is no overlap in Reynolds numbers between the two data sets. Also shown in the figure are the measurements for the glycerol-water mixture without particles; these points are correlated by $M(\phi=0)/M_{lam}=0.0123 Re_b^{0.68}$. As mentioned in context of Fig. \ref{fig3}, the pure fluid measurements are found to have transitioned to turbulent flow. The glycerol-water data for $\phi=0.1$ parallel the data set for the pure fluid; the data for $\phi=0.2$ and 0.3 also follow the pure fluid curve although the dependence on Reynolds number is not as strong. Like the pure fluid measurements, the glycerol-water data appear to have transitioned to a turbulent flow; the sharp transition is a result of the experimental design. Although not shown, the dimensionless torque measurements for the glycerol-ethanol mixture without particles was approximately $M/M_{lam} =1$ up to $Re_b=4,000$.

\begin{figure}
\centering
{\includegraphics[width=0.9\textwidth]{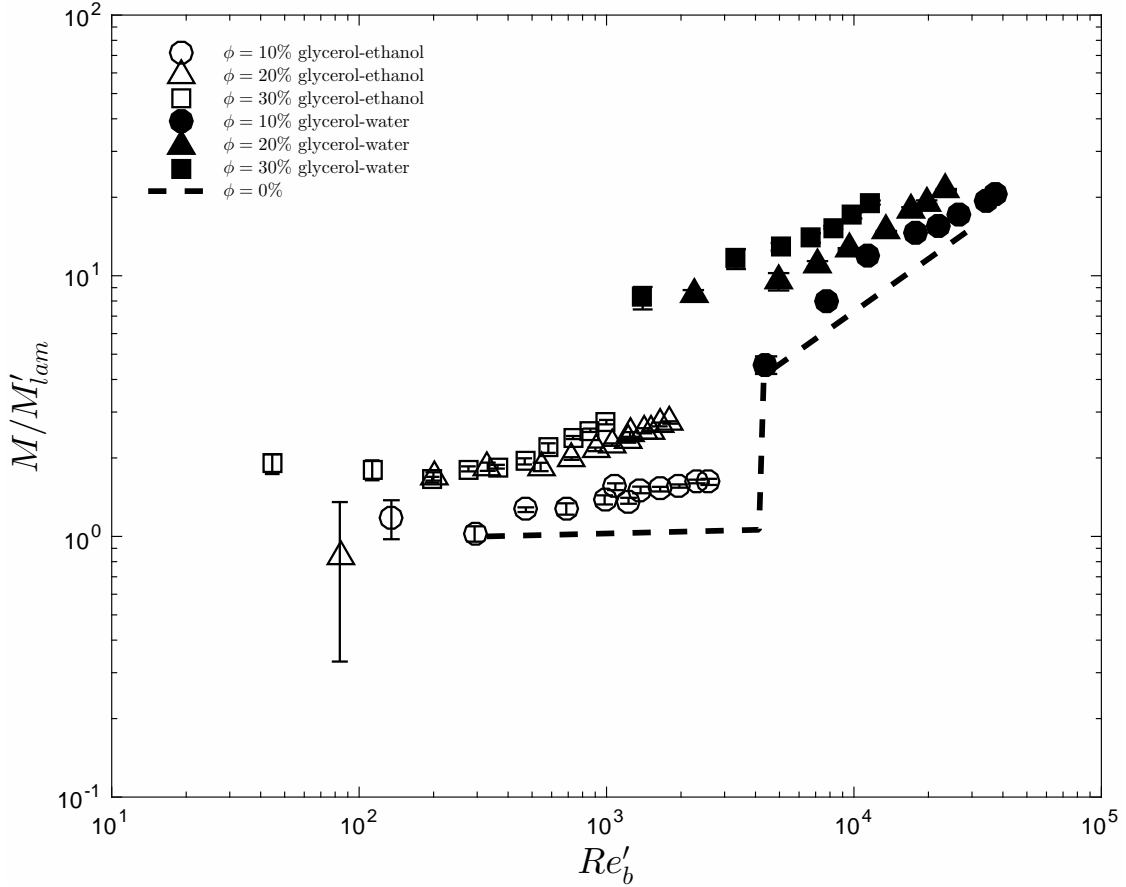}}% Images in 100% size
\caption{Measured torque normalized by effective laminar torque (using the effective viscosity from the Krieger-Dougherty relation) as a function of effective gap Reynolds number for polystyrene particles  immersed in the high- and low-viscosity fluids, respectively.}
\label{fig8}
\end{figure}

Figure \ref{fig8} presents the torque measurements $M/M_{lam}$ for both data sets for $\phi=0.1$, 0.2, and 0.3; however, with the purpose of separating the effect of solid fraction, the viscosity used in calculating $M_{lam}$ and $Re_b$ is obtained considering the Krieger-Dougherty relation $\mu'=\mu(1-\phi/\phi_m )^{-1.82} $  and using $\phi_m= 0.58$. That is:
\begin{equation}
Re_{b}'=\frac{\rho \omega r_o b}{\mu(1-\phi/\phi_m )^{-1.82}}
\end{equation}

\begin{equation}
M'_{lam}=2\pi \mu (1-\phi/\phi_m)^{-1.82}\dot{\gamma}Hr_i^2
\end{equation}

This expression was also used by \citet{Matas:2003} for transitional flows with particle Reynolds numbers beyond 10.  Also, shown by the dotted line, is the best fit to the single-phase flow measurements. Note that, as found in Fig. \ref{fig2}, the pure fluid data shows a transition to turbulent flow at $Re_b \approx4,000$. For the measurements with particles, the data suggest a similar flow transition around $Re_{b}'\approx3,000$; however, the critical Reynolds number shows dependence on solid fraction with the higher solid fractions transitioning at slightly lower Reynolds numbers. This variation in the critical Reynolds number with solid fraction is similar to that seen by \citet{Matas:2003} for flow in a pipe with relatively large particles.

\begin{figure}
\centering
\includegraphics[width=0.9\textwidth]{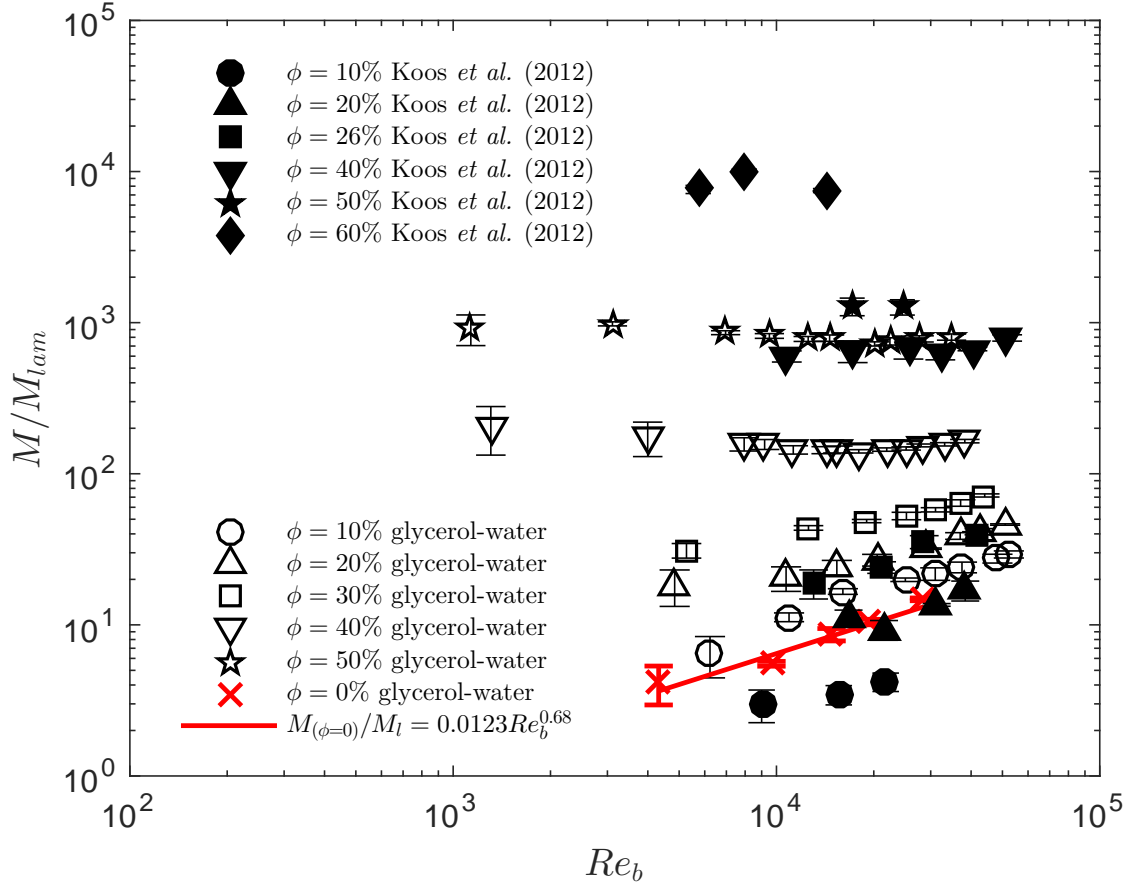}% Images in 100% size
\caption{Comparison between the normalized torques for the low-viscosity fluid and the results from \citet{Koos:2012} for rough walls experiments.}
\label{fig9}
\end{figure}

Figure \ref{fig9} shows the normalized torque measured for the glycerol-water experiments along with some data from the work by \citet{Koos:2012} involving the rough-walled rheometer. As noted earlier, the normalized torques from \citet{Koos:2012}  rough-walled experiments were independent of Reynolds number. However in reconsidering the data calibration as described in section \ref{sec2}, the normalized torque measurements for the lowest solid fractions do show a variation with Reynolds number as found in figure \ref{fig9}. For the lowest solid fractions, the \citet{Koos:2012} torque measurements are lower than in the current data. Moreover, the earlier data for $\phi=0.1$ is lower than the torque measurements for a pure fluid. This difference in magnitude is probably due to the difference in roughness between the 2012 measurements and the current data set. In 2012, the roughness was created by pressing and gluing the particles into a 1-mm rubber sheet, which created a less rough surface compared with the current experiments that used a thinner (0.15 mm) sheet. Hence the torque measurements are affected by the extent of the roughness of the surface.

\begin{figure}
\centering
\includegraphics[width=0.9\textwidth]{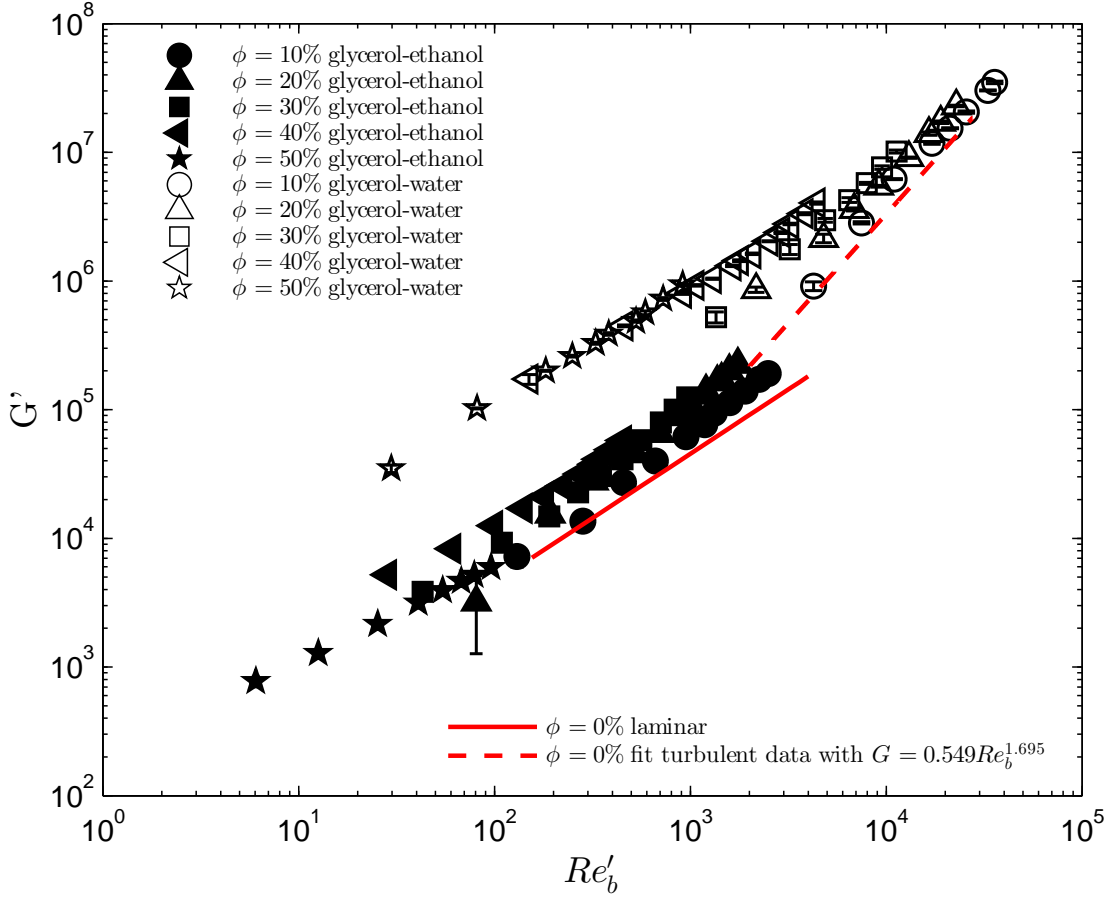}% Images in 100% size
\caption{Normalized torque measurements for all of the current experimental measurements, $G'$ as a  function of $Re_b'$ using an effective viscosity model.}
\label{fig10}
\end{figure}

In Fig. \ref{fig10}, the torque measurements for all of the experiments are presented using the normalization $G'=M/(H {\mu}'^2 / \rho)$ as a function of $Re_{b}'$ using the Krieger-Dougherty model for $\mu'$ based on the form used in \citet{Matas:2003}. The solid and dashed lines are the results for a pure fluid. As noted in the context of Eqn (2.2), the magnitude of $G'$ depends linearly on $Re_b'$ for laminar flow and was found to depend on $Re_b^{1.7}$ for the pure-fluid experiments beyond the critical Reynolds number. As shown in the figure, the data using glycerol-ethanol (high-viscosity fluid) for the lowest values of $Re_b'$ and for all solid fractions follow the laminar flow result; however, the slopes for the highest solid fractions are shallower than that for laminar flow. Around $Re_b'$ of 500, the data   begin to deviate from the laminar flow result and show a higher dependence on the Reynolds number. For the experiments using glycerol-water (low-viscosity fluid) for $\phi=0.1$, the experimental results closely match the results for a pure fluid. For $\phi=0.2$ and 0.3 the results show a weaker dependence on $Re_b'$ as compared with the pure fluid; an effective viscosity model is also less appropriate for predicting the experimental data. For $\phi=0.4$ and 0.5, the normalized torques $G'$ have a weaker dependence on Reynolds number. The data also show a difference in the magnitude of the normalized torques between the results for glycerol-ethanol and the glycerol-water mixtures; possible reasons for the differences are given in the next section. Note that the two parameters used in the Krieger-Dougherty model (the value of $\phi_m$ and the exponent) can be varied to provide a better collapse of the high-viscosity data at the lowest values of $Re_b'$. However modifications to the Krieger-Dougherty parameters do not simultaneously provide a collapse of the low-viscosity measurements.

\section{Discussion and Conclusion}\label{discussion}

For neutrally-buoyant suspension of particles in which $Re >1$, the bulk stress increases with Reynolds number due to the inertia of the particles and the fluid \citep{Batchelor:1970}. The presence of the particles also results in velocity fluctuations that enhance the transport \citep{Cartellier:2001, Martinez:2007, Mendez:2013}. As the Reynolds number is increased further, the inertia of the flow may cause a transition to turbulence. Prior studies have suggested that the particle may either enhance or suppress the turbulence \citep{Gore:1989,Matas:2003}. In addition if the particle Stokes number is greater than 10, collisions between particles or with a bounding surface may also increase the bulk stress \citep{Joseph:2001}.

In the current experiments, the measured torque is used to compute the effective viscosity. For $Re<100$ and $\phi=0.10$, the effective viscosity shows an increase with Reynolds number, which is comparable with recent numerical simulations \citep{Kulkarni:2008, Yeo:2013, Picano:2013}. According to \citet{Kulkarni:2008}, the increase results from the inertial effects on the calculation of the stresslet and is not due to the particle or velocity fluctuations. For $\phi$=0.2 and for $\phi$=0.3 for $10<Re<100$, the effective viscosity in the experiments also increases with Reynolds but at a rate that is higher than the numerical simulations. Because the experiments do not include detailed measurements of the particle or fluid velocities, it is not possible to conclude the reason for the difference. One possibility, however, may involve the roughness along the side walls, which reduces slip along the sidewalls and increases the particle velocity fluctuations \citep{Koos:2012}. As previously noted the simulations were either for a simple shear flow or for a bounded flow using smooth walls; the wall bounded simulations showed considerable wall slip, especially for the higher solid fractions \citep{Kulkarni:2008, Picano:2013}. Hence, the roughness may increase the velocity fluctuations of both the fluid and solid phases resulting in an increase in the Reynolds stresses beyond that calculated in the numerical studies. For $Re<10$, the Stokes numbers are less than 10 so that particle collisions are not expected to contribute to the stresses. Within the corresponding range of $Re_b$, the pure fluid results are laminar.

For higher values of $Re_b$, the pure fluid experiments show a transition to turbulence as shown in Figure 8; the results for the corresponding experiments with particles also show a transition. For $Re>100$ and for $\phi<0.30$, the effective viscosities show an increased dependence on the Reynolds number as compared with the results for the lower Reynolds numbers. In terms of the critical Reynolds number, the results show a slight decrease as the solid fraction is increased from $\phi=0.1$ to $\phi=0.3$.  Hence the particles not only increase the effective viscosity but also increase the fluid velocity fluctuations and promote turbulence. In addition, the torque and shear stresses are also affected by the wall roughness as found in the comparison with the data from \citet{Koos:2012}.

In looking at all of the data, as found in Fig. \ref{fig10}, the experiments for the high-viscosity fluid can be modeled using a modified version of the Krieger-Dougherty effective viscosity model. When presented in this form, there is a transition beginning around $Re_b'=500$ to a higher dependence on $Re$ that extends from the high-viscosity to the low-viscosity data. Although the results for $\phi$=0.10 are well predicted using the turbulent result with an effective viscosity, the data for the higher-solid fractions show a weaker dependence on Reynolds number. Hence the enhancement of the velocity fluctuations may not be as strong as the solid fraction is increased.

The non-dimensional torque data $\phi=0.4$ and 0.5 for the low-viscosity fluid show a roughly linearly dependence on the Reynolds number, similar to that for a laminar flow. However, the magnitude of the torques is significantly larger, and larger than predicted using the Krieger-Dougherty effective viscosity model. Earlier studies at high solid fractions suggest a change in the microstructure that shows a layering of the particles within the flow \citep{Kulkarni:2008, Yeo:2013}.  Although these studies were done at lower Reynolds numbers, similar changes in the microstructure may develop, especially for small gap size. Moreover, the differences between normalized torque for same high volume fraction but different interstitial liquid might be due to an increase in effective volume fraction due to settling, especially for the low-viscosity fluid. At the highest concentrations, a slight increase in volume fraction can result in a higher effective viscosity and a significantly lower value of $G'$.

Because the corresponding Stokes numbers for the glycerol-water experiments are from $St=10$ to $St=120$, particle collisions may also contribute to the stresses. Within this range of Stokes numbers, the coefficient of restitution increases from 0 to approximately 0.8 \citep{Joseph:2001, Yang:2006}. Hence, particle collisions may increase the stress transmission, especially for the higher Stokes number and more dilute flows.

It is important to note that all of the experiments were conducted for a single value of $b/d \approx 8$. Hence, future experiment should be performed for a range of particle sizes. In addition it would be useful to measure the velocity fluctuations and determine the microstructure of the flow to understand how the flow transitions as the solid fraction and Reynolds number increase.

\begin{acknowledgments}
We thank Prof. C.E. Brennen for his comments and fruitful discussions. R.Z. is grateful to the Fulbright-Garcia Robles foundation and to the PASPA-DGAPA-UNAM program for their financial support during his sabbatical year at Caltech.
\end{acknowledgments}

\bibliography{rheology}

\end{document}